\documentclass[english]{article}
\usepackage[T1]{fontenc}
\usepackage[latin9]{inputenc}
\usepackage{geometry}
\usepackage{float}
\usepackage{amsmath}
\usepackage{amssymb}
\usepackage{graphicx}

\makeatletter

\providecommand{\tabularnewline}{\\}

\PassOptionsToPackage{normalem}{ulem} 
\usepackage{babel}
\usepackage{cite}
\usepackage{color}
\usepackage{dsfont}
\usepackage{empheq}
\usepackage{framed}
\usepackage{mathrsfs}
\usepackage{needspace}
\usepackage{setspace}
\usepackage{tabularx}
\usepackage{tikz}
\usepackage{ulem}
\usepackage{url}

\let\@fnsymbol\@arabic

\newcommand{\mm}{\mu_{\mathrm{macro}}}
\newcommand{\lm}{\lambda_{\mathrm{macro}}}
\newcommand{\mh}{\mu_{\mathrm{micro}}}
\newcommand{\lh}{\lambda_{\mathrm{micro}}}
\newcommand{\me}{\mu_{e}}
\newcommand{\mc}{\mu_{c}}
\newcommand{\lle}{\lambda_{e}}

\newcommand{\mLc}{\me L_{c}^{2}}

\newcommand{\nablau}{\,\nabla u\,}
\newcommand{\p}{{P}}
\newcommand{\nablap}{\nabla \p}
\newcommand{\Curl}{\,\mathrm{Curl}}
\newcommand{\dev}{\, \mathrm{dev}}
\newcommand{\Div}{\mathrm{Div}}
\newcommand{\tr}{\, \mathrm{tr}}
\newcommand{\sym}{\, \mathrm{sym}\,}

\renewcommand{\skew}{\, \mathrm{skew}\,}

\renewcommand{\skew}{\, \mathrm{skew}}

\newcommand{\id}{\,\mathds{1}}

\definecolor{Green}{rgb}{0,0.52,0}

\newcommand*{\widefbox}[1]{\fbox{\hspace{2em}#1\hspace{2em}}}

\title{\vspace{-1.0cm}On the role of micro-inertia in enriched continuum mechanics}

\author{
Angela Madeo\footnote{Angela Madeo, corresponding author, angela.madeo@insa-lyon.fr, LGCIE, INSA-Lyon, Université de Lyon, 20 avenue	Albert Einstein, 69621, Villeurbanne cedex and IUF, Institut universitaire de France, 1 rue Descartes, 75231 Paris Cedex 05, France}\, and  
Patrizio Neff\,\footnote{Patrizio Neff, patrizio.neff@uni-due.de, Head of Chair for 	Nonlinear Analysis and Modelling, Fakultät für Mathematik, Universität Duisburg-Essen,  Mathematik-Carrée, Thea-Leymann-Straße 9, 45127 Essen, Germany}\, and
Elias C. Aifantis\footnote{	Elias C. Aifantis, mom@mom.gen.auth.gr, Aristotle University of Thessaloniki, Thessaloniki, 54124, Greece, Michigan Technological University, Houghton, MI 49931, USA, ITMO University, St. Petersburg, 197101, Russia} 
 \, and 
 Gabriele Barbagallo\footnote{Gabriele Barbagallo, gabriele.barbagallo@insa-lyon.fr, LaMCoS-CNRS \& LGCIE, INSA-Lyon, Universitité de Lyon, 20 avenue Albert Einstein, 69621, Villeurbanne cedex, France} \\\ and 
Marco Valerio 	d'Agostino\footnote{Marco Valerio d'Agostino, marco-valerio.dagostino@insa-lyon.fr, LGCIE, INSA-Lyon, Université de Lyon, 20 avenue Albert Einstein, 69621, Villeurbanne cedex, France} }

\makeatother

\usepackage{babel}
\begin{document}
\maketitle \addtocounter{footnote}{5} \vspace{0.5cm}
 
\begin{abstract}
In this paper the role of gradient micro-inertia terms $\overline{\eta}\lVert\nablau_{,t}\rVert^{2}$
and free micro-inertia terms $\eta\lVert\p_{,t}\rVert^{2}$ is investigated
to unveil their respective effect on the dynamical behavior of band-gap
metamaterials. We show that the term $\overline{\eta}\lVert\nablau_{,t}\rVert^{2}$
alone is only able to disclose relatively simplified dispersive behaviors.
On the other hand, the term $\eta\lVert\p_{,t}\rVert^{2}$ is in charge
of the description of the full complex behavior of band-gap metamaterials.
A suitable mixing of the two micro-inertia terms allows to describe
a new feature of the relaxed-micromorphic model, i.e. the description
of a second band-gap occurring for higher frequencies. We also show
that a split of the gradient micro-inertia $\overline{\eta}\lVert\nablau_{,t}\rVert^{2}$,
in the sense of Cartan-Lie decomposition of matrices, allows to flatten
separately longitudinal and transverse optic branches thus giving
the possibility of a second band-gap. Finally, we investigate the
effect of the gradient inertia $\overline{\eta}\lVert\nablau_{,t}\rVert^{2}$
on more classical enriched models as the Mindlin-Eringen and the internal
variable ones. We find that the addition of such gradient micro-inertia
allows for the onset of one band-gap in the Mindlin-Eringen model
and of three band-gaps in the internal variable model. In this last
case, however, non-local effects cannot be accounted for which is
a too drastic simplification for most metamaterials. We conclude that,
even when adding gradient micro-inertia terms, the relaxed micromorphic
model remains the most performing one, among the considered enriched
model, for the description of non-local band-gap metamaterials. 
\end{abstract}
\vspace{1cm}

\hspace{-0.55cm}\textbf{Keywords}: gradient micro-inertia, free micro-inertia,
complete band-gaps, non-local effects, relaxed micromorphic model,
generalized continuum models, multi-scale modeling

\vspace{1cm}

\hspace{-0.55cm}\textbf{AMS 2010 subject classification}: 74A10 (stress),
74A30 (nonsimple materials), 74A60 (micromechanical theories), 74E15
(crystalline structure), 74M25 (micromechanics), 74Q15 (effective
constitutive equations)

\pagebreak{}

\tableofcontents{}\vspace{1.2cm}

\section{Introduction}

The question of effectively studying the dynamical behavior of microscopically
heterogeneous materials in the simplified framework of continuum mechanics
is a major challenge for engineering sciences.

Indeed, it is rather clear at the present state of knowledge that
classical Cauchy continuum models are too simplified to describe the
behavior of a huge class of materials in the dynamical regime. As
a matter of fact, almost all real materials show dispersive behaviors
with respect to wave propagation, especially when considering waves
with small wavelengths (higher frequencies). More precisely, this
means that the speed of propagation of waves is not a constant, as
it happens for Caychy continua, but it depends on the wavelength of
the traveling wave. Such phenomenon is rather comprehensible if one
thinks to the fact that the mechanical properties of materials vary
when going down to lower scales. It is then sensible that the speed
of propagation of mechanical waves varies when considering waves with
wavelength which are small enough to be comparable to the characteristic
size of the underlying heterogeneities.

If Cauchy continuum theories are not rich enough to catch these dispersive
behaviors, generalized continuum theories have disclosed the possibility
of describing wave dispersion while still remaining in the framework
of continuum mechanics. Although various generalized continuum models
have been introduced to describe dispersion (see, among others, the
pioneering works \cite{eringen1964nonlinear,mindlin1964micro}), it
is yet not completely clear whether such dispersive properties can
be attributed to the constitutive assumptions which are made on the
strain energy density or to the choice of the micro-inertia terms
which can be introduced.

The aforementioned considerations about the dispersive behavior of
materials can be reformulated with renewed awareness when talking
about metamaterials.

Metamaterials are man-made artifacts which are conceived by assembling
small structural elements in periodic or quasi-periodic patterns in
such a way that the resulting material shows new astonishing mechanical
properties. The characteristic size of microstructures in such metamaterials
is much higher than the characteristic size of heterogeneities in
more classical materials. In fact, metamaterials' microstructures
usually have characteristic sizes ranging from microns to centimeters,
so that it is not necessary to go down to the molecular or the atomic
scale to be aware of their discreteness. It is thus not astonishing
that metamaterials start showing dispersive behaviors for wavelengths
which are much higher than those needed to unveil dispersion in classical
materials.

More than this, some metamaterials can exhibit dynamical behaviors
which are by far more complex than the simple dispersion. For example,
some metamaterials are able to inhibit wave propagation within certain
frequency ranges due to the presence of an underlying microstucture
which is able to resonate locally when excited at those frequencies
or even to remain completely unperturbed. The energy of the incident
wave remains trapped at the level of the microstructure and the macroscopic
propagation results to be inhibited \cite{armenise2010phononic,spadoni2009phononic,steurer2007photonic,man2013photonic}.

In order to catch the complex behavior exhibited by such metamaterials
while remaining in the framework of continuum mechanics, generalized
continuum models with enriched kinematics are needed. This means that
extra degrees of freedom must be introduced in the spirit of micromorphic
theories \cite{eringen1964nonlinear,mindlin1964micro} which allow
to account for micro-motions at the level of the microstructure. More
particularly, the kinematical unknowns of such micromorphic models
are usually the macro-displacements $u$ and the micro-distortion
tensor $\p$. Well adapted constitutive choices must then be introduced
for the strain energy density in order to well describe the behavior
of the considered metamaterials in the static regime.

As a last point, the inertia of the considered continuum must be introduced
to model its mechanical behavior in the dynamic regime. It is exactly
this point that will be the focus of the present paper: how to choose
well-suited micro-inertia terms when dealing with enriched continuum
models? How each of these terms affects the dynamic behavior of real
band-gap metamaterials? Some hints on the role of micro-inertia to
model dispersive behaviors are given in \cite{askes2011gradient}
but many fundamental questions still remain open.

We will show in this paper that: 
\begin{itemize}
\item Gradient micro-inertia terms $\overline{\eta}\left\Vert \nablau_{,t}\right\Vert ^{2}$
only allow to describe dispersion either in classical or enriched
continuum models \cite{askes2011gradient} ,
\item Micro-inertia terms involving time derivatives of the extra kinematical
degrees of freedom $\eta\left\Vert \p_{,t}\right\Vert ^{2}$ allow
to describe and control optic branches in the dispersion relations
of classical and relaxed micromorphic continuum models \cite{eringen1964nonlinear,mindlin1964micro,madeo2016reflection,madeo2015wave,neff2015relaxed,madeo2014band,madeo2016first,neff2014unifying,madeo2016complete,ghiba2014relaxed}, 
\item The relaxed micromorphic model with micro-inertia of the type $\eta\left\Vert \p_{,t}\right\Vert ^{2}$
is able to describe the onset of the first band-gaps in mechanical
metamaterials \cite{madeo2016reflection,madeo2015wave,madeo2014band,madeo2016complete,madeo2016first}, 
\item The relaxed micromorphic model with both micro-inertia terms $\eta\left\Vert \p_{,t}\right\Vert ^{2}$
and $\overline{\eta}\left\Vert \nablau_{,t}\right\Vert ^{2}$ allows
to account for the first and also for the second band-gap which occurs
for higher frequencies,
\item Classical Mindlin-Eringen models with full micro-inertia $\eta\left\Vert \p_{,t}\right\Vert ^{2}$
and $\overline{\eta}\left\Vert \nablau_{,t}\right\Vert ^{2}$ allow
for the description of only the first band-gap. 
\item Internal variable models with full micro-inertia $\eta\left\Vert \p_{,t}\right\Vert ^{2}$
and $\overline{\eta}\left\Vert \nablau_{,t}\right\Vert ^{2}$ allow
for the description of three band-gaps, even if some peculiar phenomena
related to non-locality cannot be accounted for and the behavior thus
results to be too simplified to model realistic metamaterials. 
\end{itemize}
Finally, we show that a weighted gradient micro-inertia of the type
$\frac{1}{2}\overline{\eta}_{1}\left\Vert \dev\sym\nablau_{,t}\right\Vert ^{2}+\frac{1}{2}\overline{\eta}_{2}\left\Vert \skew\nablau_{,t}\right\Vert ^{2}+\frac{1}{6}\overline{\eta}_{3}\tr\left(\nablau_{,t}\right)^{2}$
allows to flatten some optic curves independently for longitudinal
and transverse waves. More precisely, if the parameter $\overline{\eta}_{3}$
allows to flatten one optic curve for longitudinal waves, the parameter
$\overline{\eta}_{2}$ has an analogous effect for transverse waves.
Such improved control on the dispersion curves will allow for a more
effective fitting procedure on real band-gap metamaterials, since
the description of the second band-gap occurring at higher frequencies
becomes accessible.

\section{The relaxed micromorphic model}

Our novel relaxed micromorphic model endows Mindlin-Eringen's representation
with the second order \textbf{dislocation density tensor} $\alpha=-\Curl\p$
instead of the full gradient $\nablap$.\footnote{The dislocation tensor is defined as $\alpha_{ij}=-\left(\Curl\p\right)_{ij}=-\p_{ih,k}\epsilon_{jkh}$,
where $\epsilon$ is the Levi-Civita tensor and Einstein notation
of sum over repeated indices is used.} In the isotropic case the elastic energy reads

\begin{align}
W= & \underbrace{\me\,\lVert\sym\left(\nablau-\p\right)\rVert^{2}+\frac{\lle}{2}\left(\mathrm{tr}\left(\nablau-\p\right)\right)^{2}}_{\mathrm{{\textstyle isotropic\ elastic-energy}}}+\hspace{-0.1cm}\underbrace{\mc\,\lVert\skew\left(\nablau-\p\right)\rVert^{2}}_{\mathrm{{\textstyle rotational\ elastic\ coupling}}}\hspace{-0.1cm}\label{eq:Ener-2}\\
 & \quad+\underbrace{\mh\,\lVert\sym\p\rVert^{2}+\frac{\lh}{2}\,\left(\mathrm{tr}\p\right)^{2}}_{\mathrm{{\textstyle micro-self-energy}}}+\hspace{-0.2cm}\underbrace{\frac{\mLc}{2}\,\lVert\Curl\p\rVert^{2}}_{\mathrm{{\textstyle isotropic\ curvature}}}\,,\nonumber 
\end{align}
where the parameters and the elastic stress are analogous to the standard
Mindlin-Eringen micromorphic model. The model is well-posed in the
statical and dynamical case including when $\mc=0$, see \cite{neff2015relaxed,ghiba2014relaxed}.

In our relaxed model the complexity of the general micromorphic model
has been decisively reduced featuring basically only symmetric gradient
micro-like variables and the $\Curl$ of the micro-distortion $\p$.
However, the relaxed model is still general enough to include the
full micro-stretch as well as the full Cosserat micro-polar model,
see \cite{neff2014unifying}. Furthermore, well-posedness results
for the statical and dynamical cases have been provided in \cite{neff2014unifying}
making decisive use of recently established new coercive inequalities,
generalizing Korn's inequality to incompatible tensor fields \cite{neff2015poincare,neff2012maxwell,neff2011canonical,bauer2014new,bauer2016dev}.

The relaxed micromorphic model counts 6 constitutive parameters in
the isotropic case ($\me$, $\lle$, $\mh$, $\lh$, $\mc$, $L_{c}$).
The characteristic length $L_{c}$ is intrinsically related to non-local
effects due to the fact that it weights a suitable combination of
first order space derivatives in the strain energy density \eqref{eq:Ener-2}.
For a general presentation of the features of the relaxed micromorphic
model in the anisotropic setting, we refer to \cite{barbagallo2016transparent}.

As for the kinetic energy, we consider in this paper that it takes
the following form: 
\begin{gather}
J=\hspace{-0.1cm}\underbrace{\frac{1}{2}\rho\left\Vert u_{,t}\right\Vert ^{2}}_{\text{Cauchy inertia}}+\hspace{-0.1cm}\underbrace{\frac{1}{2}\eta\left\Vert \p_{,t}\right\Vert ^{2}}_{\text{free micro-inertia}}\hspace{-0.3cm}+\hspace{0.1cm}\underbrace{\frac{1}{2}\overline{\eta}_{1}\left\Vert \dev\sym\nablau_{,t}\right\Vert ^{2}+\frac{1}{2}\overline{\eta}_{2}\left\Vert \skew\nablau_{,t}\right\Vert ^{2}+\frac{1}{6}\overline{\eta}_{3}\tr\left(\nablau_{,t}\right)^{2}}_{\text{new gradient micro-inertia}},\label{eq:Kinetic}
\end{gather}
where $\rho$ is the value of the average macroscopic mass density
of the considered metamaterial, $\eta$ is the free micro-inertia
density and the $\overline{\eta}_{i},i=\{1,2,3\}$ are the gradient
micro-inertia densities associated to the different terms of the Cartan-Lie
decomposition of $\nablau$.

The associated equations of motion in strong form, obtained by a classical
least action principle take the form (see \cite{madeo2016reflection,madeo2015wave,neff2015relaxed,madeo2014band})
\begin{align}
\rho\,u_{,tt}+\hspace{-0,8cm}\underbrace{\Div[\,\mathcal{I}\,]}_{\text{new augmented term}}\hspace{-0,8cm} & =\Div\left[\,\widetilde{\sigma}\,\right], & \eta\,\p_{,tt} & =\widetilde{\sigma}-s-\Curl\,m,\label{eq:Dyn}
\end{align}
where 
\begin{align}
\mathcal{I} & =\overline{\eta}_{1}\,\dev\sym\nablau_{,tt}+\overline{\eta}_{2}\,\skew\nablau_{,tt}+\frac{1}{3}\overline{\eta}_{3}\tr\left(\nablau_{,tt}\right),\nonumber \\
\widetilde{\sigma} & =2\,\me\,\sym\left(\nablau-\p\right)+\lle\,\tr\left(\nablau-\p\right)\id+2\,\mc\,\skew\left(\nablau-\p\right),\\
s & =2\,\mh\,\sym\p+\lh\,\tr\left(\!\p\right)\id,\nonumber \\
m & =\mLc\,\Curl\p.\nonumber 
\end{align}
The fact of adding a gradient micro-inertia in the kinetic energy
\eqref{eq:Kinetic} modifies the strong-form PDEs of the relaxed micromorphic
model with the addition of the new term $\mathcal{I}$. Of course,
boundary conditions would also be modified with respect to the ones
presented in \cite{madeo2016reflection,madeo2016first}. The study
of the new boundary conditions induced by gradient micro-inertia will
be the object of a subsequent paper where the effect of such extra
terms on the conservation of energy will also be analyzed.

\section{Plane wave propagation}

Sufficiently far from a source, dynamic wave solutions may be treated
as planar waves. Therefore, we now want to study harmonic solutions
traveling in an infinite domain for the differential system \eqref{eq:Dyn}.
We suppose that the space dependence of all introduced kinematical
fields are limited to the scalar component $X$ which is also the
direction of propagation of the wave. To do so, following \cite{madeo2014band,madeo2015wave,madeo2016first,madeo2016reflection,madeo2016complete,neff2016real}
we define: 
\begin{align}
\p^{S} & :=\frac{1}{3}\tr\left(\p\right), & \p_{\left[ij\right]} & :=\left(\skew\p\right)_{ij}=\frac{1}{2}\left(\p_{ij}-\p_{ji}\right),\label{Decom}\\
\p^{D} & :=\p_{11}-\p^{S}, & \p_{\left(ij\right)} & :=\left(\sym\p\right)_{ij}=\frac{1}{2}\left(\p_{ij}+\p_{ji}\right),\nonumber \\
P^{V} & :=P_{22}-P_{33}.\nonumber 
\end{align}
With this decomposition, equations \eqref{eq:Dyn} can be rewritten
as (see \cite{madeo2014band,madeo2015wave}) 
\begin{itemize}
\item a set of three equations only involving longitudinal quantities: 
\begin{align}
\rho\,\ddot{u}_{1}-\hspace{-0.2cm}\underbrace{\frac{2\,\overline{\eta}_{1}+\overline{\eta}_{3}}{3}\,\ddot{u}_{1,11}}_{\text{new augmented terms}}\hspace{-0.2cm} & =\left(2\,\me+\lle\right)u_{1,11}-2\me\,P_{,1}^{D}-(2\mu_{e}+3\lambda_{e})\,P_{,1}^{S}\,,\vspace{0.4cm}\nonumber \\
\eta\,\ddot{P}^{D} & =\frac{4}{3}\,\me\,u_{1,1}+\frac{1}{3}\,\mLc\,P_{,11}^{D}-\frac{2}{3}\,\mLc P_{,11}^{S}-2\left(\me+\mh\right)\,P^{D}\,,\vspace{0.4cm}\label{Long}\\
\eta\,\ddot{P}^{S} & =\frac{2\,\me+3\,\lle}{3}\,u_{1,1}-\frac{1}{3}\,\mLc P_{,11}^{D}+\frac{2}{3}\,\mLc P_{,11}^{S}\nonumber \\
 & \quad-\left(2\,\me+3\,\lle+2\,\mh+3\,\lh\right)\,P^{S}\,,\nonumber 
\end{align}

\item two sets of three equations only involving transverse quantities in
the $\xi$-th direction, with $\xi=2,3$: 
\begin{align}
\rho\,\ddot{u}_{\xi}-\hspace{-0.3cm}\underbrace{\frac{\,\overline{\eta}_{1}+\overline{\eta}_{2}}{2}\,\ddot{u}_{\xi,11}}_{\text{new augmented terms}}\hspace{-0.3cm} & =\left(\me+\mc\right)u_{\xi,11}-2\,\me\,P_{\left(1\xi\right),1}+2\,\mc\,P_{\left[1\xi\right],1},\vspace{0.4cm}\nonumber \\
\eta\,\ddot{P}_{\left(1\xi\right)} & =\me\,u_{\xi,1}+\frac{1}{2}\,\mLc\,P_{(1\xi)}{}_{,11}+\frac{1}{2}\,\mLc\,P_{\left[1\xi\right],11}\label{Trans}\\
 & \quad-2\left(\me+\mh\right)\,P_{(1\xi)},\vspace{0.4cm}\nonumber \\
\eta\,\ddot{P}_{\left[1\xi\right]} & =-\mc\,u_{\xi,1}+\frac{1}{2}\,\mLc\,P_{(1\xi),11}+\frac{1}{2}\,\mLc P_{\left[1\xi\right]}{}_{,11}-2\,\mc\,P_{\left[1\xi\right]},\nonumber 
\end{align}

\end{itemize}

\begin{itemize}
\item One equation only involving the variable $P_{\left(23\right)}$: 
\begin{align}
\eta\,\ddot{P}_{\left(23\right)}=-2\left(\me+\mh\right)P_{\left(23\right)}+\mLc P_{\left(23\right),11},\label{Shear}
\end{align}

\item One equation only involving the variable $P_{\left[23\right]}$ :
\begin{align}
\eta\,\ddot{P}_{\left[23\right]}=-2\,\mc\,P_{\left[23\right]}+\mLc P_{\left[23\right],11},\label{Rotations23}
\end{align}

\item One equation only involving the variable $P^{V}$: 
\begin{align}
\eta\,\ddot{P}^{V}=-2\left(\me+\mh\right)P^{V}+\mLc P_{,11}^{V}.\label{VolumeVariation}
\end{align}

\end{itemize}
Once this symplified system of PEDEs is obtained, we look for a wave
form solution of the type: 
\begin{equation}
\underbrace{\mathbf{v}_{1}(X,t)=\boldsymbol{\beta}\,e^{i(kX-\omega t)}}_{\text{longitudinal}},\qquad\underbrace{\mathbf{v}_{\tau}(X,t)=\boldsymbol{\gamma}\,^{\tau}e^{i(kX-\omega t)}}_{\text{transversal}},\quad\tau=2,3,\qquad\underbrace{\mathbf{v}_{4}(X,t)=\boldsymbol{\gamma}\,^{4}e^{i(kX-\omega t)}}_{\text{uncoupled}},\label{WaveForm2}
\end{equation}
where we set for compactness 
\begin{align}
\mathbf{v}_{1}=\left(u_{1},P^{D},P^{S}\right)\qquad\mathbf{v}_{\tau}=\left(u_{\tau},P_{(1\tau)},P_{[1\tau]}\right),\quad\tau=2,3,\qquad\mathbf{v}_{4}=\left(P_{(23)},P_{[23]},P^{V}\right).
\end{align}
where $\boldsymbol{\beta}=(\beta_{1},\beta_{2},\beta_{3})^{T}\in\mathbb{C}^{3}$,
$\boldsymbol{\gamma}^{\tau}=(\gamma_{1}^{\tau},\gamma_{2}^{\tau},\gamma_{3}^{\tau})^{T}\in\mathbb{C}^{3}$
and $\boldsymbol{\gamma}^{4}=(\gamma_{1}^{4},\gamma_{2}^{4},\gamma_{3}^{4})^{T}\in\mathbb{C}^{3}$
are the unknown amplitudes of the considered waves\footnote{Here, we understand that having found the (in general, complex) solutions
of \eqref{WaveForm2} only the real or imaginary parts separately
constitute actual wave solutions which can be observed in reality.}, $k$ is the wavenumber and $\omega$ is the wave-frequency.

Replacing the wave form solution \eqref{WaveForm2} in Eqs. \eqref{Long},
\eqref{Trans}, \eqref{Shear}, \eqref{Rotations23} and \eqref{VolumeVariation},
it is possible to express the system as: 
\begin{equation}
\mathbf{A}_{1}\cdot\boldsymbol{\beta}=0,\qquad\qquad\mathbf{A}_{\tau}\cdot\boldsymbol{\gamma}^{\tau}=0,\quad\tau=2,3,\qquad\qquad\mathbf{A}_{4}\cdot\boldsymbol{\gamma}^{4}=0,\label{AlgSys}
\end{equation}
where 
\begin{align}
\mathbf{A}_{1}(\omega,k)\, & =\left(\begin{array}{ccc}
-\omega^{2}\left(1+k^{2}\,\frac{2\,\overline{\eta}_{1}+\overline{\eta}_{3}}{3\,\rho}\right)+c_{p}^{2}\,k^{2} & \,i\:k\,2\,\me/\rho\  & i\:k\:\left(2\,\me+3\,\lle\right)/\rho\\
\\
-i\:k\,\frac{4}{3}\,\me/\eta & -\omega^{2}+\frac{1}{3}k^{2}c_{m}^{2}+\omega_{s}^{2} & -\frac{2}{3}\,k^{2}c_{m}^{2}\\
\\
-\frac{1}{3}\,i\,k\:\left(2\,\me+3\,\lle\right)/\eta & -\frac{1}{3}\,k^{2}\,c_{m}^{2} & -\omega^{2}+\frac{2}{3}\,k^{2}\,c_{m}^{2}+\omega_{p}^{2}
\end{array}\right),\nonumber \\
\nonumber \\
\mathbf{A}_{2}(\omega,k)=\mathbf{A}_{3}(\omega,k)\, & =\left(\begin{array}{ccc}
-\omega^{2}\left(1+k^{2}\,\frac{\overline{\eta}_{1}+\overline{\eta}_{2}}{2\,\rho}\right)+k^{2}c_{s}^{2}\  & \,i\,k\,2\,\me/\rho\  & -i\,\frac{k}{\rho}\eta\,\omega_{r}^{2},\\
\\
-\,i\,k\,\me/\eta, & -\omega^{2}+\frac{c_{m}^{2}}{2}k^{2}+\omega_{s}^{2} & \frac{c_{m}^{2}}{2}k^{2}\\
\\
\frac{i}{2}\,\omega_{r}^{2}\,k & \frac{c_{m}^{2}}{2}k^{2} & -\omega^{2}+\frac{c_{m}^{2}}{2}k^{2}+\omega_{r}^{2}
\end{array}\right),\nonumber \\
\\
\mathbf{A}_{4}(\omega,k)\, & =\left(\begin{array}{ccc}
-\omega^{2}+c_{m}^{2}\,k^{2}+\omega_{s}^{2} & 0 & 0\\
\\
0 & -\omega^{2}+c_{m}^{2}\,k^{2}+\omega_{r}^{2} & 0\\
\\
0 & 0 & -\omega^{2}+c_{m}^{2}\,k^{2}+\omega_{s}^{2}
\end{array}\right).\nonumber 
\end{align}
Here, we have defined:

\begin{empheq}[box=\widefbox]{align} c_{m}&=\sqrt{\frac{\mLc}{\eta}},\qquad& c_{s}&=\sqrt{\frac{\me+\mc}{\rho}},\qquad &c_{p}&=\sqrt{\frac{2\,\me+\lle}{\rho}},\nonumber \\ \nonumber \\ \omega_{s}&=\sqrt{\frac{2\left(\me+\mh\right)}{\eta}},\qquad &\omega_{p}&=\sqrt{\frac{\left(2\,\me+3\,\lle\right)+\left(2\,\mh+3\,\lh\right)}{\eta}},\qquad&\omega_{r}&=\sqrt{\frac{2\,\mc}{\eta}},\nonumber
\end{empheq}

In order to have non-trivial solutions of the algebraic systems (\ref{AlgSys}),
one must impose that 
\begin{equation}
\underbrace{\mathrm{det}\,\mathbf{A}_{1}(\omega,k)=0,}_{\text{longitudinal}}\qquad\qquad\underbrace{\mathrm{det}\,\mathbf{A}_{2}(\omega,k)=\mathrm{det}\,\mathbf{A}_{3}(\omega,k)=0,}_{\text{transverse}}\qquad\qquad\underbrace{\mathrm{det}\,\mathbf{A}_{4}(\omega,k)=0,}_{\text{uncoupled}}\label{Dispersion}
\end{equation}
The solutions $\omega=\omega(k)$ of these algebraic equations are
called the dispersion curves of the relaxed micromorphic model for
longitudinal, transverse and uncoupled waves, respectively.

In what follows we will present the results obtained for the numerical
values of the elastic coefficients chosen as in Table \ref{ParametersValues}
if not differently specified. 
\begin{table}[H]
\begin{centering}
\begin{tabular}{|c|c|c|}
\hline 
Parameter  & Value  & Unit\tabularnewline
\hline 
\hline 
$\me$  & $200$  & MPa\tabularnewline
\hline 
$\lle=2\me$  & 400  & MPa\tabularnewline
\hline 
$\mc=5\me$  & 1000  & MPa\tabularnewline
\hline 
$\mh$  & 100  & MPa\tabularnewline
\hline 
$\lh$  & $100$  & MPa\tabularnewline
\hline 
$L_{c}\ $  & $1$  & mm\tabularnewline
\hline 
$\rho$  & $2000$  & kg/m$^{3}$\tabularnewline
\hline 
\end{tabular}\quad{}\quad{}\quad{}\quad{}%
\begin{tabular}{|c|c|c|}
\hline 
Parameter  & Value  & Unit\tabularnewline
\hline 
\hline 
$\lm$  & $82.5$  & MPa\tabularnewline
\hline 
$\mm$  & $66.7$  & MPa\tabularnewline
\hline 
$E_{\mathrm{macro}}$  & $170$  & MPa\tabularnewline
\hline 
$\nu_{\mathrm{macro}}$  & $0.28$  & $-$\tabularnewline
\hline 
\end{tabular}
\par\end{centering}

\caption{\label{ParametersValues}Values of the parameters used in the numerical
simulations (left) and corresponding values of the Lamé parameters
and of the Young modulus and Poisson ratio (right), for the formulas
needed to calculate the homogenized macroscopic parameters starting
from the microscopic ones, see \cite{barbagallo2016transparent}.}
\end{table}

In the following sections we will explicitly discuss which is the
effect of each micro-inertia parameter on the dispersion curves of
the relaxed micromorphic model. More particularly, we will focus on
the cases: 
\begin{itemize}
\item vanishing free micro-inertia $\eta=0$ and non-vanishing gradient
micro-inertia, 
\item both non-vanishing gradient micro-inertia and free micro-inertia. 
\end{itemize}
The remaining case (vanishing gradient micro-inertia $\overline{\eta}=0$
and non-vanishing free micro-inertia $\eta\neq0$) is the classical
case treated for the relaxed micromorphic model in \cite{madeo2014band,madeo2015wave,madeo2016first,madeo2016reflection,madeo2016complete,neff2016real}.
To the sake of completeness, we present here again the dispersion
curves for this case when using the values of the parameters given
in Tab. \ref{ParametersValues}. 
\begin{figure}[H]
\begin{centering}
\includegraphics[height=6.2cm]{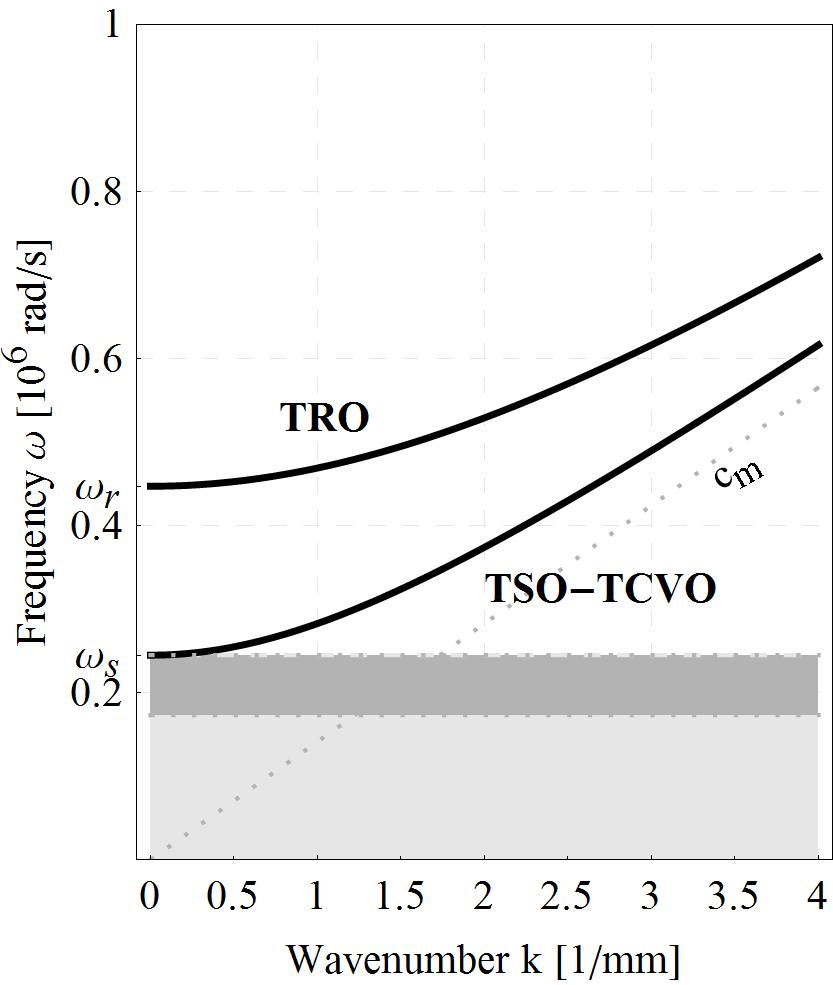} \includegraphics[height=6.2cm]{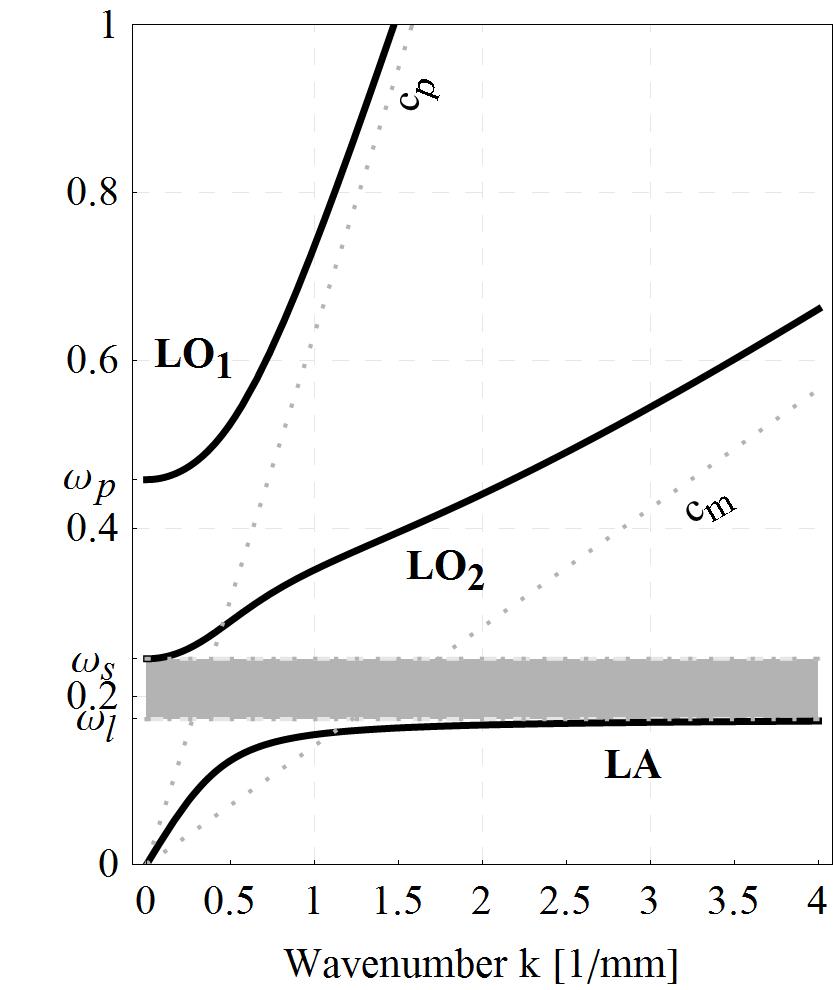}
\includegraphics[height=6.2cm]{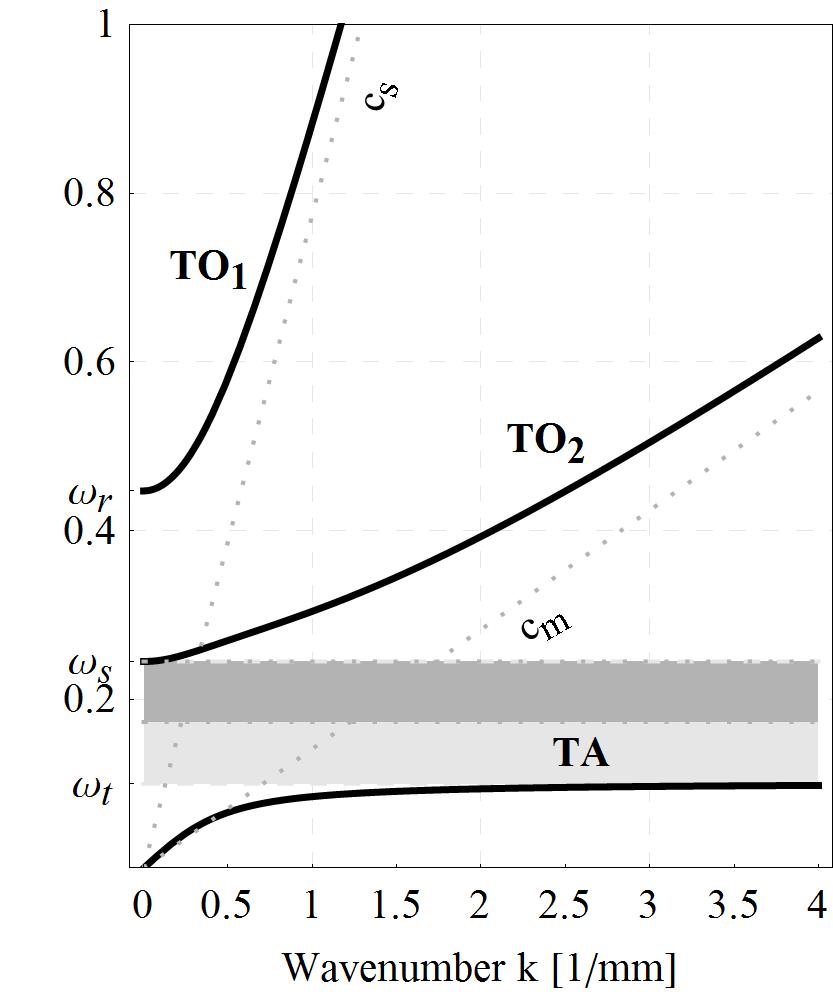} 
\par\end{centering}

\caption{Dispersion relations $\omega=\omega(k)$ for the uncoupled (left),
longitudinal (center) and transverse (right) waves of the \textbf{relaxed
micromorphic model} with free micro-inertia $\eta=10^{-2}kg/m$. \label{Relaxed} }
\end{figure}

It can be found that, when considering the free micro-inertia alone,
the relaxed micromorphic model is able to predict the first band gap
which usually occurs at relatively low frequencies. Moreover, the
relaxed micromorphic model is, to the current state of the art, the
only continuum model which is able to describe simultaneously band-gaps
and non-local behavior \cite{madeo2016first}.

In the next sections we will present the new results concerning the
effect of the gradient micro-inertia terms on the dispersion curves
of the relaxed micromorphic model, as well as the effect of such gradient
micro-inertia on more classical enriched models (Mindlin, internal
variable).

\section{Case of vanishing free micro-inertia $\eta$ and non-vanishing gradient
micro-inertia $\overline{\eta}$}

In this section we discuss the effect, on the dispersion curves of
enriched continuum models, of the gradient micro-inertia term alone.
We will show that the fact of complementing the macro-inertia $\rho\lVert u_{,t}\rVert^{2}$
only with the gradient micro-inertia $\overline{\eta}\lVert\nablau_{,t}\rVert^{2}$
is a fundamental modeling limitation since the complexity of the dynamical
behavior of micromorphic models cannot be unveiled. Nevertheless,
the gradient micro-inertia allows to describe some dispersion which
is not allowed by classical Cauchy models.

\subsection{Study of the dispersion curves}

\begin{figure}[H]
\begin{centering}
\includegraphics[width=6cm]{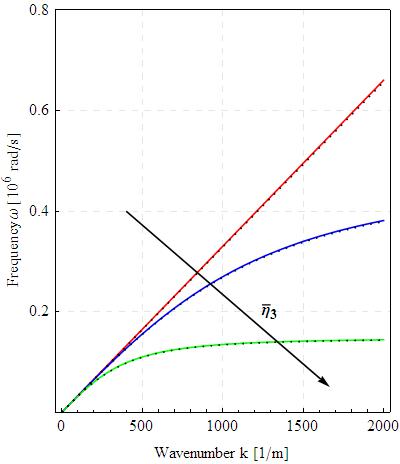} \hspace{2cm}
\includegraphics[width=6cm]{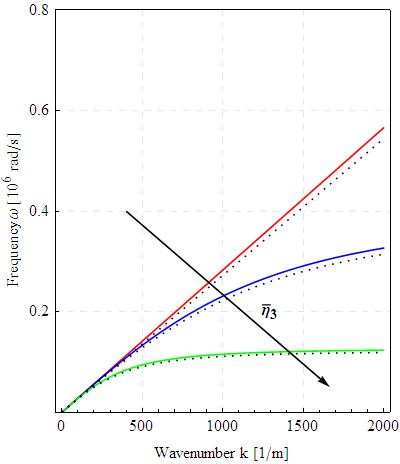} 
\par\end{centering}

\caption{\label{EtaBarLong}Dispersion relations $\omega=\omega(k)$ for the
longitudinal waves of the \textbf{relaxed micromorphic model} with
gradient micro-inertia $\overline{\eta}_{3}=(0,3\times10^{-3},3\times10^{-2})kg/m$
and $\eta=0$. Dotted in black the dispersion relations for a \textbf{first
gradient model} with Lamé parameters $\mm$ and $\lm$ and the same
inertiae $\rho$ and $\overline{\eta}_{3}$ (left). The same picture
obtained imposing $\lh=0$ (right): a very slight variation with respect
to the $1^{st}$ gradient case can be detected. }
\end{figure}

In the case in which we consider only the gradient micro-inertia $\overline{\eta}\neq0$
to be non-vanishing, the matrix associated to the longitudinal dynamical
system can be expressed as\footnote{We can notice from the form of $\mathbf{A}_{1}(\omega,k)$ that considering
an additional micro-inertia $\overline{\eta}$ is equivalent to defining
an average macroscopic density depending on the wavelength as $\rho^{*}(k)=\rho+k^{2}\,\overline{\eta}$.
The same can be found for the transverse waves.}: 
\begin{align}
\mathbf{A}_{1}(\omega,k)\, & =\left(\begin{array}{ccc}
-\omega^{2}\left(\rho+k^{2}\,\frac{2\,\overline{\eta}_{1}+\overline{\eta}_{3}}{3}\right)+(2\,\me+\lle)\,k^{2} & \,i\:k\:2\me\  & i\:k\:\left(2\,\me+3\,\lle\right)\\
\\
-i\:k\,\frac{4}{3}\,\me & \frac{1}{3}k^{2}\mLc+2\,(\me+\mh) & -\frac{2}{3}\,k^{2}\mLc\\
\\
-\frac{1}{3}\,i\,k\:\left(2\,\me+3\,\lle\right) & -\frac{1}{3}\,k^{2}\,\mLc & \frac{2}{3}\,k^{2}\,\mLc+\omega_{p}^{2}
\end{array}\right).
\end{align}
It is possible to remark that the polynomial $\mathrm{det}\,\mathbf{A}_{1}(\omega,k)$
is of the second order in $\omega$. This implies that we have a unique
positive solution of the equation $\mathrm{det}\,\mathbf{A}_{1}(\omega,k)=0$
when considering positive $k$ \footnote{It can be checked that, when considering elastic parameters which
guarantee positive definiteness of the elastic energy the solution
$\omega=\omega(k)$ of the characteristic polynomials are always real
\cite{neff2016real}.}. In particular, when plotting such solution in the $(\omega,k)$
plane only one acoustic branch can be detected (see Fig.$\ $\ref{EtaBarLong})\footnote{Here and in the sequel, we will always set $\overline{\eta}_{1}=0$,
since we could not isolate a characteristic effect of such parameters
on the dispersion curves.}. 

Comparing the results shown in Fig.$\ $\ref{EtaBarLong} with those
presented in Fig.$\ $\ref{Relaxed}, it is immediate to notice that
the fact of considering the gradient micro-inertia alone significantly
constrains the behavior of the considered enriched continuum. Even
if the constitutive expression for the strain energy density $W$
is the same both in Fig.$\ $\ref{EtaBarLong} and in Fig.$\ $\ref{Relaxed}
(see Eq. \eqref{eq:Ener-2}), the fact of using a gradient micro-inertia
$\overline{\eta}\lVert\nablau_{,t}\rVert^{2}$ instead of a free micro-inertia
$\eta\lVert\p_{,t}\rVert^{2}$ drastically simplifies the patterns
which are found for the dispersion curves. With reference to Fig.$\ $\ref{EtaBarLong},
we can remark that a unique acoustic wave is found and that the presence
of a non-vanishing micro-inertia parameter $\overline{\eta}_{3}$
induces a dispersive behavior. When the gradient micro-inertia parameters
are all vanishing ($\overline{\eta}_{1}=\overline{\eta}_{2}=\overline{\eta}_{3}=0$),
this means that only a macro-inertia $\rho\lVert u_{,t}\rVert^{2}$
is present and this correspond to an almost constant speed of the
traveling waves, which it is what happens for the classical Cauchy
case. It can be shown that, considering an adapted choice of the constitutive
parameters for the relaxed micromorphic model with macro-inertia $\rho\lVert u_{,t}\rVert^{2}$
alone, the dispersion curve obtained is exactly the straight one obtained
with classical Cauchy model.

\medskip{}

\begin{figure}[H]
\begin{centering}
\includegraphics[width=6cm]{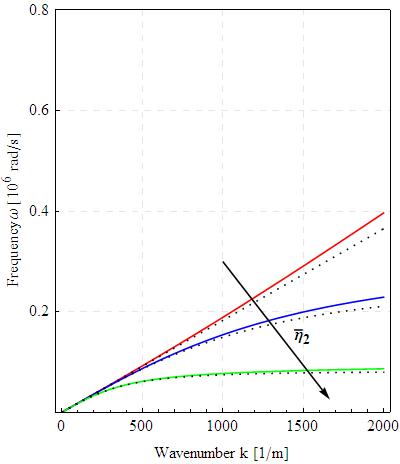} 
\par\end{centering}

\caption{\label{EtaBarTran}Dispersion relations $\omega=\omega(k)$ for the
transverse waves of the \textbf{relaxed micromorphic model} with gradient
micro-inertia $\overline{\eta}_{2}=(0,2\times10^{-3},2\times10^{-2})kg/m$
and $\eta=0$ and dotted in black the dispersion relations for a \textbf{first
gradient model} with Lamé parameters $\mm$ and $\lm$ and the same
inertiae $\rho$ and $\overline{\eta}_{t}$. }
\end{figure}

With a similar reasoning as the one made for longitudinal waves, considering
the case $\overline{\eta}\neq0$ for transverse waves, the matrix
associated to the transverse dynamical system can be expressed as
\begin{align}
\mathbf{A}_{2}(\omega,k)= & \left(\begin{array}{ccc}
-\omega^{2}\left(\rho+k^{2}\,\frac{\overline{\eta}_{1}+\overline{\eta}_{2}}{2}\right)+k^{2}(\me+\mc) & \,i\,k\,2\me\  & -i\,k\,2\mc\\
\\
-i\,k\,2\me & \mLc k^{2}+4(\me+\mh) & \mLc k^{2}\\
\\
i\,k\,2\mc & \mLc k^{2} & \mLc\,k^{2}+4\mc
\end{array}\right),
\end{align}
It is possible to see that the new inertia terms $\overline{\eta}_{2}$
plays the same role for the transverse waves that was played by $\overline{\eta}_{3}$
for the longitudinal waves. The results concerning the solutions $\omega=\omega(k)$
of the characteristic equation $\mathrm{det}\,\mathbf{A}_{2}(\omega,k)=0$
are analogous to the case of longitudinal waves, see Fig.$\ $\ref{EtaBarTran}.

\medskip{}

If the particular case with non-null gradient micro-inertia $\overline{\eta}\neq0$
and null free micro-inertia $\eta=0$ is considered, the matrix associate
to the uncoupled waves reduces to: 
\begin{align}
\mathbf{A}_{4}(\omega,k)\, & =\left(\begin{array}{ccc}
\mLc\,k^{2}+2\left(\me+\mh\right) & 0 & 0\\
\\
0 & \mLc\,k^{2}+2\,\mc & 0\\
\\
0 & 0 & \mLc\,k^{2}+2\left(\me+\mh\right)
\end{array}\right).
\end{align}
from which it is not possible to derive any dispersion curve, due
to the absence of inertia terms.

\subsection{A first conclusion on the effect of gradient micro-inertia on enriched
continuum models.}
\begin{itemize}
\item When considering a macro-inertia term $\rho\lVert u_{,t}\rVert^{2}$
alone, only one acoustic wave is present and the associated dispersion
has an almost constant speed of propagation. Such behavior is strongly
dictated by the macro-inertia term since the difference on the associated
dispersion curves between a simple Cauchy energy $W(\nablau)$ and
an enriched model $W=W(\nablau,\p,\Curl\,\p)$ is small and vanishing
considering an adapted choice of the constitutive parameters. 
\item When considering a complementation of the macro-inertia $\rho\lVert u_{,t}\rVert^{2}$
with a gradient micro-inertia $\overline{\eta}\lVert\nablau_{,t}\rVert^{2}$
the speed of propagation of waves is not constant anymore, but it
depends on the wavelength of the traveling waves. Nevertheless, only
an acoustic branch can be described, independently of the more or
less complicated (standard or enriched) kinematics. 
\item Complementing the macro-inertia $\rho\lVert u_{,t}\rVert^{2}$ with
a free micro-inertia $\eta\lVert\p_{,t}\rVert^{2}$ allows to disclose
the full rich constitutive behavior provided by the fact of considering
an enriched model, as studied in \cite{madeo2014band,madeo2015wave,madeo2016first,madeo2016reflection,madeo2016complete,neff2016real}
and reproduced in Fig.$\,$\ref{Relaxed}. Two optic waves are observed,
both for longitudinal and transverse waves, in addition to the acoustic
ones already discussed in the previous case (see Fig.$\ $\ref{Relaxed}).
The properties of such curves depend both on the constitutive parameters
appearing in the expression of the energy (Eq. \eqref{eq:Ener-2})
and on the free inertia parameter $\eta$. In this framework of inertia
of the type $\rho\lVert u_{,t}\rVert^{2}+\eta\lVert\p_{,t}\rVert^{2}$
, the relaxed micromorphic model is the only non-local, enriched continuum
model allowing for the presence of band-gaps \cite{madeo2016complete}. 
\end{itemize}

\section{Case of both non-vanishing free micro-inertia $\eta$ and gradient
micro-inertia $\overline{\eta}$}

In this section we will discuss the effect of a full inertia $\rho\lVert u_{,t}\rVert^{2}+\eta\lVert\p_{,t}\rVert^{2}$
on the dispersion curves of the relaxed micromorphic model. We will
show that the complementation of the macro inertia with both the gradient
and free micro-inertia allows for the description of a new feature
of the relaxed micromorphic model, i.e. the onset of a second band-gap
occurring at higher frequencies with respect to the first one.

\subsection{Dispersion relations}

\begin{figure}[H]
\begin{centering}
\includegraphics[width=6cm]{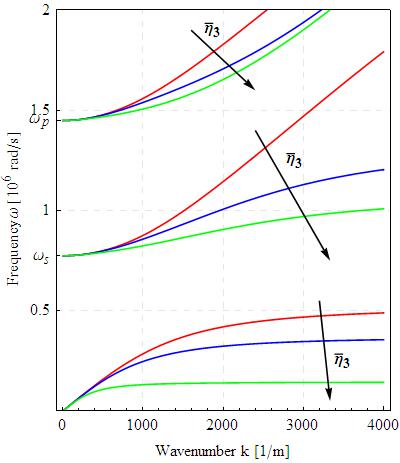} \hspace{2cm}
\includegraphics[width=6cm]{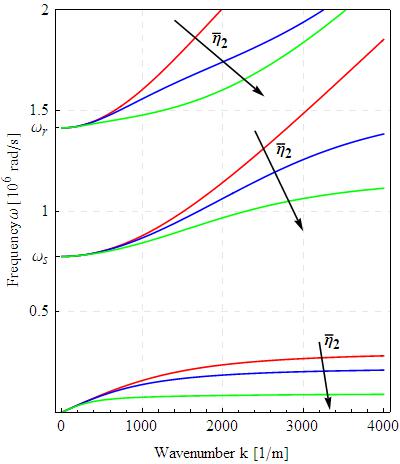} 
\par\end{centering}

\caption{\label{EtaFullLong}Dispersion relations $\omega=\omega(k)$ of the
\textbf{relaxed micromorphic model} for the longitudinal waves with
free micro-inertia $\eta=10^{-3}$ and gradient micro-inertia $\overline{\eta}_{3}=(3\times10^{-4},3\times10^{-3},3\times10^{-2})kg/m$
(left) and transverse waves with micro-inertia $\eta=10^{-3}$ and
gradient micro-inertia $\overline{\eta}_{2}=(2\times10^{-4},2\times10^{-3},2\times10^{-2})kg/m$
(right). }
\end{figure}

Now, we show in Figure \ref{EtaFullLong} the results obtained for
non-null micro-inertia $\eta\neq0$ with the addition of gradient
micro-inertia $\overline{\eta}\neq0$. Surprisingly, the combined
effect of the traditional micro-inertia $\eta$ with the gradient
micro-inertiae can lead to the onset of a second longitudinal and
transverse band gap. Moreover, it is possible to notice that the addition
of gradient micro-inertiae $\overline{\eta}_{1}$, $\overline{\eta}_{2}$
and $\overline{\eta}_{3}$ has no effect on the cut-off frequencies,
which only depend on the free micro-inertia $\eta$ (and of course
on the constitutive parameters). 

The uncoupled waves in the relaxed micromorphic model with generalized
inertia behave as in the relaxed micromorphic model as it is possible
to see analyzing the matrix: 
\begin{align}
\mathbf{A}_{4}(\omega,k)\, & =\left(\begin{array}{ccc}
-\omega^{2}+c_{m}^{2}\,k^{2}+\omega_{s}^{2} & 0 & 0\\
\\
0 & -\omega^{2}+c_{m}^{2}\,k^{2}+\omega_{r}^{2} & 0\\
\\
0 & 0 & -\omega^{2}+c_{m}^{2}\,k^{2}+\omega_{s}^{2}
\end{array}\right).
\end{align}
The resulting dispersion curves are the same to the ones obtained
with the classical relaxed micromorphic model, see Fig.$\ $\ref{Relaxed},
right.

\subsection{Cut-offs and asymptotic behavior}

To study the asymptotic behavior of the dispersion curves for the
relaxed micromorphic model with full inertia, let us introduce the
following quantities:

	\begin{empheq}[box=\widefbox]{align} \omega_{v}&=\sqrt{\frac{\left(2\,\me+\lle\right)+\left(2\,\mh+\lh\right)}{\eta}},&\omega_{l}&=\sqrt{\frac{2\,\mh+\lh}{\eta}},\qquad&\omega_{t}&=\sqrt{\frac{\mh}{\eta}},\nonumber \\ \nonumber \\\nonumber 	\omega_{\overline{l}}&=\sqrt{\frac{2\,\me+\lle}{\frac{2\,\overline{\eta}_1+\overline{\eta}_3}{3}}}, &\qquad\omega_{\overline{t}}&=\sqrt{\frac{2\,(\mc+\me)}{\frac{\overline{\eta}_1+\overline{\eta}_2}{2}}}.\hspace{-0.5cm} 	\end{empheq}

As stated in the previous section the cut-off frequencies are not
modified by the insertion of a gradient micro-inertia term. Therefore,
considering the longitudinal waves, we have one acoustic branch of
the dispersion curve and two optic branches with cut-off frequencies:
\begin{align}
\omega_{s}= & \sqrt{\frac{2\left(\me+\mh\right)}{\eta}}, & \omega_{p}= & \sqrt{\frac{\left(2\,\me+3\,\lle\right)+\left(2\,\mh+3\,\lh\right)}{\eta}},
\end{align}
On the other hand, the asymptotic behavior changes in a radical fashion
from the classical relaxed micromorphic model. The horizontal asymptote
of the acoustic curve changes and we have the onset of a new horizontal
asymptote for one of the optic branches, which values are respectively:
\begin{align}
\omega_{l,\mathrm{acoustic}}= & \sqrt{\frac{\omega_{\overline{l}}^{2}+\omega_{v}^{2}-\sqrt{(\omega_{\overline{l}}^{2}+\omega_{v}^{2})^{2}-4\,\omega_{\overline{l}}^{2}\,\omega_{l}^{2}}}{2}},\\
\omega_{l,\mathrm{optic}}= & \sqrt{\frac{\omega_{\overline{l}}^{2}+\omega_{v}^{2}+\sqrt{(\omega_{\overline{l}}^{2}+\omega_{v}^{2})^{2}-4\,\omega_{\overline{l}}^{2}\,\omega_{l}^{2}}}{2}}.\nonumber 
\end{align}
No difference is found in the other optic branch that has an asymptote
with slope $c_{m}$ as in the classical relaxed micromorphic model.

Analogously, considering the transverse waves, we have one acoustic
branch and two optic branches with cut-off frequencies: 
\begin{align}
\omega_{s}= & \sqrt{\frac{2\left(\me+\mh\right)}{\eta}}, & \omega_{r}= & \sqrt{\frac{2\,\mc}{\eta}}.
\end{align}
Once again, the horizontal asymptote of the acoustic curve changes
with respect to the classical relaxed case and we have an extra horizontal
asymptote for one of the optic branches, which values are respectively:
\begin{align}
\omega_{t,\mathrm{acoustic}}= & \frac{1}{2}\sqrt{\omega_{\overline{t}}^{2}+\omega_{s}^{2}+\omega_{r}^{2}-\sqrt{(\omega_{\overline{t}}^{2}+\omega_{s}^{2}+\omega_{r}^{2})^{2}-4\,\omega_{\overline{t}}^{2}\,\omega_{t}^{2}}},\\
\omega_{t,\mathrm{optic}}= & \frac{1}{2}\sqrt{\omega_{\overline{t}}^{2}+\omega_{s}^{2}+\omega_{r}^{2}+\sqrt{(\omega_{\overline{t}}^{2}+\omega_{s}^{2}+\omega_{r}^{2})^{2}-4\,\omega_{\overline{t}}^{2}\,\omega_{t}^{2}}}.\nonumber 
\end{align}
No difference is found in the other optic branch that has an asymptote
with slope $c_{m}$ as in the classical relaxed micromorphic model.

Finally, no change whatsoever is present in the uncoupled waves that
keep having cut-off frequencies $\omega_{s}$ and $\omega_{r}$ and
oblique asymptote of slope $c_{m}$.

\section{Combined effect of the free and gradient micro-inertiae on more classical
enriched models (Mindlin-Eringen and internal variable)}

In this section, we discuss the effect on the Mindlin-Eringen and
the internal variable model of the addition of the gradient micro-inertia
$\overline{\eta}\lVert\nablau_{,t}\rVert^{2}$ to the classical terms
$\rho\lVert u_{,t}\rVert^{2}+\eta\lVert\p_{,t}\rVert^{2}$. We will
show that the previously discussed effect of the parameters $\overline{\eta}_{2}$
and $\overline{\eta}_{3}$ is maintained both for the Mindlin-Eringen
and for the internal variable case.

\begin{figure}[H]
\begin{centering}
\includegraphics[width=6cm]{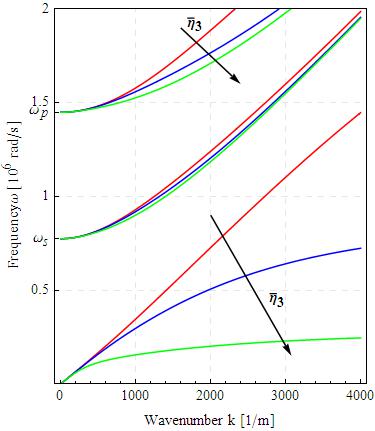} \hspace{2cm}
\includegraphics[width=6cm]{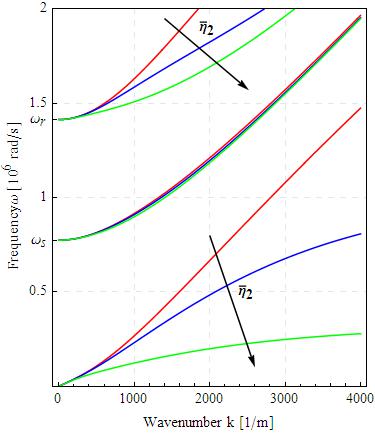} 
\par\end{centering}

\caption{\label{Mind}Dispersion relations $\omega=\omega(k)$ of the \textbf{standard
Mindlin-Eringen model} for the longitudinal waves with free micro-inertia
$\eta=10^{-3}$ and gradient micro-inertia $\overline{\eta}_{3}=(3\times10^{-4},3\times10^{-3},3\times10^{-2})kg/m$
(left) and transverse waves with micro-inertia $\eta=10^{-3}$ and
gradient micro-inertia $\overline{\eta}_{2}=(2\times10^{-4},2\times10^{-3},2\times10^{-2})kg/m$
(right). }
\end{figure}

Figure \ref{Mind} refers to the study of the effects of the parameters
$\overline{\eta}_{2}$ and $\overline{\eta}_{3}$ on the dispersion
curves of the classical Mindlin-Eringen micromorphic model. To the
sake of completeness, we recall that the (simplified) strain energy
density for this model in the isotropic case takes the form: 
\begin{align}
W= & \underbrace{\me\,\lVert\sym\left(\nablau-\p\right)\rVert^{2}+\frac{\lle}{2}\left(\mathrm{tr}\left(\nablau-\p\right)\right)^{2}}_{\mathrm{{\textstyle isotropic\ elastic-energy}}}+\hspace{-0.1cm}\underbrace{\mc\,\lVert\skew\left(\nablau-\p\right)\rVert^{2}}_{\mathrm{{\textstyle rotational\ elastic\ coupling}}}\hspace{-0.1cm}\label{eq:Ener-Mindlin}\\
 & \quad+\underbrace{\mh\,\lVert\sym\p\rVert^{2}+\frac{\lh}{2}\,\left(\mathrm{tr}\p\right)^{2}}_{\mathrm{{\textstyle micro-self-energy}}}+\hspace{-0.2cm}\underbrace{\frac{\mLc}{2}\,\lVert\nabla\,\p\rVert^{2}}_{\mathrm{{\textstyle isotropic\ curvature}}}\,,\nonumber 
\end{align}

Recalling the results of \cite{madeo2014band}, we remark that when
the gradient micro-inertia is vanishing ($\overline{\eta}_{1}=\overline{\eta}_{2}=\overline{\eta}_{3}=0$)
the Mindlin-Eringen model does not allow the description of band-gaps,
due to the presence of a straight acoustic waves. On the other hand,
when switching on the parameters $\overline{\eta}_{2}$ and $\overline{\eta}_{3}$
, the acoustic branches are flattened (they take a horizontal asymptote),
so that the first band-gap can be described. The analogous case for
the relaxed micromorphic model (Fig.$\,$\ref{Relaxed}) allowed instead
for the description of 2 band gaps.

\begin{figure}[H]
\begin{centering}
\includegraphics[width=6cm]{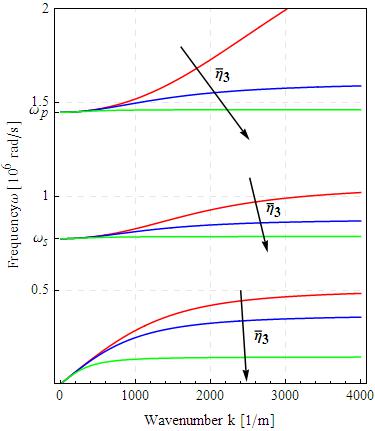} \hspace{2cm}
\includegraphics[width=6cm]{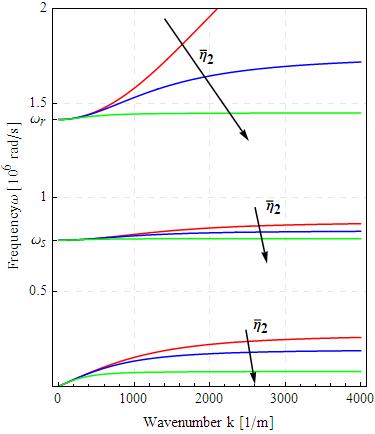} 
\par\end{centering}

\caption{\label{Int}Dispersion relations $\omega=\omega(k)$ of the \textbf{internal
variable model} for the longitudinal waves with free micro-inertia
$\eta=10^{-3}$ and gradient micro-inertia $\overline{\eta}_{3}=(3\times10^{-4},3\times10^{-3},3\times10^{-2})kg/m$
(left) and transverse waves with micro-inertia $\eta=10^{-3}$ and
gradient micro-inertia $\overline{\eta}_{2}=(2\times10^{-4},2\times10^{-3},2\times10^{-2})kg/m$
(right). }
\end{figure}

Figure \ref{Int} shows the behavior of the addition of the gradient
micro-inertia $\overline{\eta}\lVert\nablau_{,t}\rVert^{2}$ on the
internal variable model. We recall (see \cite{neff2014unifying})
that the energy for the internal variable model does not include higher
space derivatives of the micro-distortion tensor $\p$ and, in the
isotropic case, takes the form: 
\begin{align}
W= & \underbrace{\me\,\lVert\sym\left(\nablau-\p\right)\rVert^{2}+\frac{\lle}{2}\left(\mathrm{tr}\left(\nablau-\p\right)\right)^{2}}_{\mathrm{{\textstyle isotropic\ elastic-energy}}}+\hspace{-0.1cm}\underbrace{\mc\,\lVert\skew\left(\nablau-\p\right)\rVert^{2}}_{\mathrm{{\textstyle rotational\ elastic\ coupling}}}\hspace{-0.1cm}\label{eq:Ener-Int}\\
 & \quad+\underbrace{\mh\,\lVert\sym\p\rVert^{2}+\frac{\lh}{2}\,\left(\mathrm{tr}\p\right)^{2}}_{\mathrm{{\textstyle micro-self-energy}}}\,,\nonumber 
\end{align}

By direct observation of Fig.$\,$\ref{Int}, we can notice that suitably
choosing the relative position of $\omega_{r}$ and $\omega_{p}$,
the internal variable model allows to account for 3 band gaps.

We thus have an extra band-gap with respect to the analogous case
for the relaxed micromorphic model (see Fig.$\,$\ref{Relaxed}),
but we are not able to consider non-local effects. The fact of excluding
the possibility of describing non-local effects in metamaterials can
be sometimes too restrictive. For example, flattening the curve which
originates from $\omega_{r}$ and which is associated to rotational
modes of the microstructure is unphysical for the great majority of
metamaterials.

\section{Conclusions}

In this paper we discuss the fundamental role of micro-inertia in
enriched continuum models of the micromorphic type. 

We show that if, on one hand, the free micro-inertia term $\eta\lVert P_{,t}\rVert^{2}$
is strictly necessary to disclose the full rich behavior of micromorphic
media in the dynamic regime, on the other hand the gradient micro-inertia
$\overline{\eta}\lVert\nabla u_{,t}\rVert^{2}$ has the macroscopic
effect of flattening some of the dispersion curves so allowing for
the description of extra band-gaps. In particular, we show that:
\begin{itemize}
\item In the case of the relaxed micromorphic model one band-gap can be
described when introducing the free micro-inertia $\eta\lVert P_{,t}\rVert^{2}$
alone. When introducing a mixed micro-inertia $\eta\lVert P_{,t}\rVert^{2}+\overline{\eta}\lVert\nabla u_{,t}\rVert^{2}$
two band-gaps can be accounted for by the same model.
\item In the case of Mindlin-Eringen model no band-gaps are possible with
the term $\eta\lVert P_{,t}\rVert^{2}$ alone, while the onset of
a single band-gap can be granted by the addition of the extra term
$\overline{\eta}\lVert\nabla u_{,t}\rVert^{2}$.
\item In the internal variable model two band-gaps are possible with the
term $\eta\lVert P_{,t}\rVert^{2}$ alone, even if non-localities
cannot be accounted for by such model. When adding the extra term
$\overline{\eta}\lVert\nabla u_{,t}\rVert^{2}$ even three band-gaps
become possible, but the behavior of the dispersion curves becomes
fairly unrealistic for a huge class of real metamaterials.
\end{itemize}
In conclusion, the results presented in this paper confirm the preceding
findings according to wich the relaxed micromorphic model is the most
suitable enriched model for the simultaneous description of i) band-gaps
and ii) non-localities in mechanical metamaterials.

Future work will be devoted to the application of the results obtained
in this paper for the fitting of the proposed model with enriched
micro-inertia on real metamaterials exhibiting multiple band-gaps.

\section{Acknowledgments}

Angela Madeo thanks INSA-Lyon for the funding of the BQR 2016 \textquotedbl{}Caractérisation
mécanique inverse des métamatériaux: modélisation, identification
expérimentale des paramètres et évolutions possibles\textquotedbl{},
as well as the CNRS-INSIS for the funding of the PEPS project.

{\footnotesize{}\let\stdsection\section \def\section *#1{\stdsection{#1}}}{\footnotesize \par}

{\footnotesize{}\bibliographystyle{plain}
\bibliography{library}

\begin{thebibliography}{10}

\bibitem{armenise2010phononic}
Mario~N. Armenise, Carlo~E. Campanella, Caterina Ciminelli, Francesco
  Dell'Olio, and Vittorio M.~N. Passaro.
\newblock {Phononic and photonic band gap structures: Modelling and
  applications}.
\newblock {\em Physics Procedia}, 3(1):357--364, 2010.

\bibitem{askes2011gradient}
Harm Askes and Elias~C. Aifantis.
\newblock {Gradient elasticity in statics and dynamics: An overview of
  formulations, length scale identification procedures, finite element
  implementations and new results}.
\newblock {\em International Journal of Solids and Structures},
  48(13):1962--1990, 2011.

\bibitem{barbagallo2016transparent}
Gabriele Barbagallo, Marco~Valerio {d}'Agostino, Rafael Abreu, Ionel-Dumitrel
  Ghiba, Angela Madeo, and Patrizio Neff.
\newblock {Transparent anisotropy for the relaxed micromorphic model:
  macroscopic consistency conditions and long wave length asymptotics}.
\newblock {\em Preprint ArXiv}, 1601.03667, 2016.

\bibitem{bauer2014new}
Sebastian Bauer, Patrizio Neff, Dirk Pauly, and Gerhard Starke.
\newblock {New Poincar{\'{e}}-type inequalities}.
\newblock {\em Comptes Rendus Mathematique}, 352(2):163--166, 2014.

\bibitem{bauer2016dev}
Sebastian Bauer, Patrizio Neff, Dirk Pauly, and Gerhard Starke.
\newblock {Dev-Div- and DevSym-DevCurl-inequalities for incompatible square
  tensor fields with mixed boundary conditions}.
\newblock {\em ESAIM: Control, Optimisation and Calculus of Variations},
  22(1):112--133, 2016.

\bibitem{eringen1964nonlinear}
Ahmed~Cemal Eringen and Erdogan~S. Suhubi.
\newblock {Nonlinear theory of simple micro-elastic solids -- I}.
\newblock {\em International Journal of Engineering Science}, 2(2):189--203,
  1964.

\bibitem{ghiba2014relaxed}
Ionel-Dumitrel Ghiba, Patrizio Neff, Angela Madeo, Luca Placidi, and Giuseppe
  Rosi.
\newblock {The relaxed linear micromorphic continuum: existence, uniqueness and
  continuous dependence in dynamics}.
\newblock {\em Mathematics and Mechanics of Solids}, 20(10):1171--1197, 2014.

\bibitem{madeo2016first}
Angela Madeo, Gabriele Barbagallo, Marco~Valerio {d}'Agostino, Luca Placidi,
  and Patrizio Neff.
\newblock {First evidence of non-locality in real band-gap metamaterials:
  determining parameters in the relaxed micromorphic model}.
\newblock {\em Proceedings of the Royal Society A: Mathematical, Physical and
  Engineering Sciences}, 472(2190):20160169, 2016.

\bibitem{madeo2016complete}
Angela Madeo, Patrizio Neff, Marco~Valerio {d}'Agostino, and Gabriele
  Barbagallo.
\newblock {Complete band gaps including non-local effects occur only in the
  relaxed micromorphic model}.
\newblock {\em Preprint ArXiv}, 1602.04315, 2016.

\bibitem{madeo2014band}
Angela Madeo, Patrizio Neff, Ionel-Dumitrel Ghiba, Luca Placidi, and Giuseppe
  Rosi.
\newblock {Band gaps in the relaxed linear micromorphic continuum}.
\newblock {\em Zeitschrift f{\"{u}}r Angewandte Mathematik und Mechanik},
  95(9):880--887, 2014.

\bibitem{madeo2015wave}
Angela Madeo, Patrizio Neff, Ionel-Dumitrel Ghiba, Luca Placidi, and Giuseppe
  Rosi.
\newblock {Wave propagation in relaxed micromorphic continua: modeling
  metamaterials with frequency band-gaps}.
\newblock {\em Continuum Mechanics and Thermodynamics}, 27(4-5):551--570, 2015.

\bibitem{madeo2016reflection}
Angela Madeo, Patrizio Neff, Ionel-Dumitrel Ghiba, and Giussepe Rosi.
\newblock {Reflection and transmission of elastic waves in non-local band-gap
  metamaterials: a comprehensive study via the relaxed micromorphic model}.
\newblock {\em Journal of the Mechanics and Physics of Solids}, 2016.

\bibitem{man2013photonic}
Weining Man, Marian Florescu, Kazue Matsuyama, Polin Yadak, Geev Nahal, Seyed
  Hashemizad, Eric Williamson, Paul Steinhardt, Salvatore Torquato, and Paul
  Chaikin.
\newblock {Photonic band gap in isotropic hyperuniform disordered solids with
  low dielectric contrast.}
\newblock {\em Optics Express}, 21(17):19972--81, 2013.

\bibitem{mindlin1964micro}
Raymond~David Mindlin.
\newblock {Micro-structure in linear elasticity}.
\newblock {\em Archive for Rational Mechanics and Analysis}, 16(1):51--78,
  1964.

\bibitem{neff2015relaxed}
Patrizio Neff, Ionel-Dumitrel Ghiba, Markus Lazar, and Angela Madeo.
\newblock {The relaxed linear micromorphic continuum: well-posedness of the
  static problem and relations to the gauge theory of dislocations}.
\newblock {\em The Quarterly Journal of Mechanics and Applied Mathematics},
  68(1):53--84, 2014.

\bibitem{neff2014unifying}
Patrizio Neff, Ionel-Dumitrel Ghiba, Angela Madeo, Luca Placidi, and Giuseppe
  Rosi.
\newblock {A unifying perspective: the relaxed linear micromorphic continuum}.
\newblock {\em Continuum Mechanics and Thermodynamics}, 26(5):639--681, 2014.

\bibitem{neff2016real}
Patrizio Neff, Angela Madeo, Gabriele Barbagallo, Marco~Valerio {d}'Agostino,
  Rafael Abreu, and Ionel-Dumitrel Ghiba.
\newblock {Real wave propagation in the isotropic relaxed micromorphic model}.
\newblock {\em Preprint ArXiv}, 1605.07902, 2016.

\bibitem{neff2011canonical}
Patrizio Neff, Dirk Pauly, and Karl-Josef Witsch.
\newblock {A canonical extension of Korn's first inequality to H(Curl)
  motivated by gradient plasticity with plastic spin}.
\newblock {\em Comptes Rendus Mathematique}, 349(23):1251--1254, 2011.

\bibitem{neff2012maxwell}
Patrizio Neff, Dirk Pauly, and Karl-Josef Witsch.
\newblock {Maxwell meets Korn: A new coercive inequality for tensor fields in
  RNxN with square-integrable exterior derivative}.
\newblock {\em Mathematical Methods in the Applied Sciences}, 35(1):65--71,
  2012.

\bibitem{neff2015poincare}
Patrizio Neff, Dirk Pauly, and Karl-Josef Witsch.
\newblock {Poincar{\'{e}} meets Korn via Maxwell: Extending Korn's first
  inequality to incompatible tensor fields}.
\newblock {\em Journal of Differential Equations}, 258(4):1267--1302, 2015.

\bibitem{spadoni2009phononic}
Alessandro Spadoni, Massimo Ruzzene, Stefano Gonella, and Fabrizio Scarpa.
\newblock {Phononic properties of hexagonal chiral lattices}.
\newblock {\em Wave Motion}, 46(7):435--450, 2009.

\bibitem{steurer2007photonic}
Walter Steurer and Daniel Sutter-Widmer.
\newblock {Photonic and phononic quasicrystals}.
\newblock {\em Journal of Physics D: Applied Physics}, 40(13):229--247, 2007.

\end{thebibliography}
 \bibliographystyle{plain}}{\footnotesize \par}

{\footnotesize{}\let\section\stdsection}{\footnotesize \par}
\end{document}